\documentclass[12pt]{article}
\usepackage{amssymb}
\usepackage{epsfig}

\newcommand{\bq}{\begin{equation}}
\newcommand{\eq}{\end{equation}}
\newcommand{\bqa}{\begin{eqnarray}}
\newcommand{\eqa}{\end{eqnarray}}
\newcommand{\ben}{\begin{enumerate}}
\newcommand{\een}{\end{enumerate}}
\newcommand{\bc}{\begin{center}}
\newcommand{\ec}{\end{center}}

\def\gsim{\gtrsim}
\def\lsim{\lesssim}

\def\pr#1#2#3{ Phys. Rev. ${\bf{#1}}$ (#2) #3}

\def\pl#1#2#3{ Phys. Lett. ${\bf{#1}}$ (#2) #3}

\def\np#1#2#3{ Nucl. Phys. ${\bf{#1}}$ (#2) #3}
\def\zp#1#2#3{ Z. f. Phys. ${\bf{#1}}$ (#2) #3}

\def\etal{{\it et.al.\/}}

\global\nulldelimiterspace = 0pt

\def\ol#1{\overline{#1}}

\def\L{ {\cal L }}
\def\O{ {\cal O }}
\def\A{ {\cal A }}

\def\sw{s_W}
\def\cw{c_W}
\def\swd{s^2_W}
\def\cwd{c^2_W}

\def\mz{M_Z}
\def\mzd{M_Z^2}

\def\lamNP{\Lambda_{NP}}

\def\vec#1{\overrightarrow{#1}} 
\def\vtil{\tilde{v}}

\begin{document}
\thispagestyle{empty}
\begin {flushleft}
 THES-TP 97/02 \\

March 1997\\
\end{flushleft}

\vspace*{2cm}

\hspace*{-0.5cm}
\begin{center}
{\Large {\bf  Using $e^-e^+ \to b\bar b$ to test properties of new
interactions}}\\
\vspace{0.2cm}
{\Large {\bf at LEP2 and higher energies}}\footnote{Partially 
supported by the EC contract CHRX-CT94-0579.}\hspace{2.2cm}\null 
\hspace*{-0.5cm}
\vspace{1.cm} \\{\large G.J.
Gounaris$^a$, D.T. Papadamou$^a$ and F.M. Renard$^b$}\hspace{2.2cm}\null
\vspace {0.5cm} \\
$^a$Department of Theoretical Physics, University of Thessaloniki,
\hspace{2.2cm}\null \\
Gr-54006, Thessaloniki, Greece.\\ \vspace{0.2cm} $^b$ Physique
Math\'{e}matique et Th\'{e}orique, UPRES-A 5032\hspace{2.2cm}\null\\
Universit\'{e} Montpellier
II,  F-34095 Montpellier Cedex 5.\hspace{2.2cm}\null\\[1cm]

\vspace*{3cm}
{\bf Abstract}\hspace{2.2cm}\null
\end{center}
\hspace*{-1.2cm}
\begin{minipage}[b]{16cm}
We show that in $e^-e^+$ colliders at energies above  
the $Z$-peak, the process $e^-e^+ \to b\bar b$
becomes very sensitive to the presence of residual New
Physics (NP) effects described by the 
$dim=6$ $SU(3)\times SU(2)\times U(1)$ gauge invariant 
operators $\O_{qW}$, $\O_{qB}$ and $\O_{bB}$. This observation
should be combined with the already known great sensitivity
of the light fermion  production through $e^-e^+$ annihilation
above the $Z$-peak, 
to the bosonic operators $\ol{\O}_{DW}$ and $\ol{\O}_{DB}$. 
It is important to emphasize that the effects of all these 
operators are largely hidden at the $Z$-peak; 
while they are enhanced above it through the use
the  "$Z$-peak subtracted representation". 
The observability limits for detecting these operators at LEP2
and NLC, through such light fermion production processes, are also 
established.
\end{minipage} 

\setcounter{footnote}{0} 
\clearpage
\newpage 
  
\hoffset=-1.46truecm
\voffset=-2.8truecm
\textwidth 16cm
\textheight 22cm
\setlength{\topmargin}{1.5cm}

\section{Introduction.} 

It is very reasonable to expect that the part of the 
New Physics (NP) beyond the Standard Model (SM),
which is responsible for the mass generation mechanism, should 
predominantly affect
the scalar sector as well as the sectors most closely related to
it; namely the one of the heavy quarks of the 
third family and the sector of the  gauge bosons.\par

Assuming that the particles responsible for NP are too heavy to
be directly produced in the present and forthcoming Colliders, and 
integrating them out, we end up with a description of the 
residual NP effects in terms of $SU(3)\times
SU(2)\times U(1)$ gauge invariant operators constructed in terms
of the fields of the aforementioned  sectors. 
If we also assume that the 
NP scale is sufficiently
larger than the electroweak scale, then one can 
restrict the list of such
operators to those of $dimension = 6$ \cite{Buch}.
In the present discussion we will
further assume NP to be CP-invariant.\par

For purely bosonic CP conserving interactions, the NP effects 
are then described 
in terms of 11  $dim=6$ operators involving $\gamma$, 
$W$, $Z$ and Higgs boson fields \cite{Hag, DeR},
and another 3 ones involving also the gluon field \cite{topdyn}. 
Direct ways to test the presence of such interactions 
have been proposed by mainly studying boson pair 
($W^+W^-$, $HZ$, $H\gamma$) production in the $e^+e^-$ or
$\gamma\gamma$ collisions and also at hadron colliders,
\cite{LEP2, LC, hadcol}. 
Indirect tests
through fermion pair production at Z peak and beyond have been
done by looking at gauge boson self-energy corrections and also
at the vertex corrections induced by
these operators \cite{Hag1, clean, cleanlong}.\par

Related studies concerning the NP induced operators 
involving the
quarks of the third family, which are 34 in number,
have also been recently done 
\cite{Peccei, topop, Kuroda, topdyn, whisnant, Yuan}.\par

In the study of these operators, one often divides them into 
two categories, using the concepts of
"blind" and "non-blind". "Non-blind" 
operators are defined as those inducing tree-level contributions 
to observables measurable at LEP1/SLC. Such observables may include  
pair production of any lepton or quark pair, (except the top), 
as well as studies of the gauge boson 
propagators properties, that  are strongly constrained by the 
present  LEP1/SLC precision measurements. 
It is commonly believed that 
the present bounds on "non-blind" operators are very strong, and
that they cannot be substantially improved by higher energy
experiments at LEP2 or bigger Colliders.\par

It turns out that
this conclusion is erroneous for the bosonic operators
\cite{Hag1, clean, cleanlong}
\bqa
\ol{\O}_{DB} & = & 2~(\partial_{\mu}B^{\mu \rho})(\partial^\nu B_{\nu
\rho}) \ \ \  , \ \   \label{listDB} \\[0.1cm] 
\overline{\O}_{DW} & =& 2 ~ (D_{\mu} \overrightarrow W^{\mu
\rho}) (D^{\nu} \overrightarrow W_{\nu \rho})  \ \ \
  . \ \  \label{listDW} 
\eqa
Because, these operators  give tree level contributions
to the  fermion production amplitudes that are 
strongly rising with $q^2$, so that the constraints achievable
at LEP2 and higher energies should be  
substantially stronger than those possible at LEP1/SLC. 
Concerning the comparison of these operators, to the operators
\bq
\O_{DW}  =  ~ (D_{\mu} \overrightarrow W_{\nu
\rho}).(D^{\mu} \overrightarrow W^{\nu \rho})  \ \ \ , 
\ \ \ \O_{DB}  =  ~ (D_{\mu} B_{\nu \rho}) (D^{\mu} B^{\nu \rho}) 
   \ \  ,  \label{listDWDB} 
\eq
defined in \cite{Hag}, we remark that 
\bq
\O_{DW}~=~ \ol{\O}_{DW} +12 g ~ \O_W \ \ \ \ , \ \  \O_{DB}~=~
\ol{\O}_{DB} \ , \   \label{listDWDB1} 
\eq
where \cite{topdyn}
\bq
\O_W =  {1\over3!}\left( \overrightarrow{W}^{\ \ \nu}_\mu\times
  \overrightarrow{W}^{\ \ \lambda}_\nu \right) \cdot
  \overrightarrow{W}^{\ \ \mu}_\lambda \ \ \
 .  \ \  \label{listW} 
\eq
Thus, the operators $\ol{\O}_{DW}~,~\O_{DW}$ give identical tree
level contributions to the gauge boson self-energies; and the
same is true for the operators $\ol{\O}_{DB}~,~\O_{DB}$. \par

A contribution to the fermion production amplitudes which
is strongly rising with $q^2$, is interesting in two aspects. 
On the one hand it is   obviously
favouring observation at higher energy colliders. And  on the other
hand, the use of the so-called "$Z$-peak subtracted
representation", allows a clean disentangling of operators
inducing such $q^2$ dependent contributions \cite{Zsub}. 
This procedure, which consists 
in using as inputs $Z$-peak measurements and in subtracting 
NP contributions at $q^2=\mzd$, manages to express all  observables
beyond the $Z$-peak in terms of only the aforementioned specific operators, 
while all other possible contributions  
automatically cancel.\par

In \cite{clean, cleanlong} it was observed that 
$\ol{\O}_{DW}$ and $\ol{\O}_{DB}$ are the only purely bosonic
such operators, and the process $e^-e^+ \to f \bar f$, 
for any light fermion $f$, was used to study them.
In the present work we 
present a corresponding study for the NP operators 
involving heavy quarks of the third family.
Among them \cite{topdyn}, we find that only
\bqa
\O_{qW}& =& \frac{1}{2} \left (\bar q_L \gamma_{\mu}
\overrightarrow\tau q_L \right ) \cdot
(D_\nu \overrightarrow W^{\mu\nu}) \ \ \ , 
\label{listqW}\\[0.1cm]
\O_{qB} & = & \bar q_L \gamma^\mu q_L (\partial^\nu B_{\mu\nu}) \ \
\ , \label{listqB}\\[0.1cm]
\O_{bB} & = & \bar b_R \gamma^\mu b_R (\partial^\nu B_{\mu\nu}) \ \
\ , \label{listbB}
\eqa  
generate at tree level such a strong $q^2$ dependent
contribution. In
(\ref{listqW}-\ref{listbB}), $q_L=(t_L, b_L)$ is
the third family left-handed quark doublet and $D_\nu$ is the usual
covariant derivative.\par

As stated already, the application of the 
"$Z$-peak subtracted representation" to the
processes $e^+e^-\to f\bar f$, uses as inputs the experimental
values for the partial widths $\Gamma(Z\to f\bar f)$ and the 
$Z$-peak asymmetry
factors $A_f$. Under these conditions, it can been shown 
that any further NP contribution 
to $e^+e^-\to f\bar f$ must come from (some of) the five operators
 $\overline{\O}_{DW}$, $\overline{\O}_{DB}$, $\O_{qW}$, 
$\O_{qB}$ and $\O_{bB}$ only.
The two bosonic ones contribute to all fermion pair production in a
universal way, while ($\O_{qW}$, $\O_{qB}$, $\O_{bB}$)
 only contribute to
$e^+e^-\to b\bar b$ (and of course to $e^+e^-\to t\bar t$ to
which we are not interested here since 
there is no definite $Z$-peak subtraction \cite{Kuroda}). 
Since  $\overline{\O}_{DB}$ and $\overline{\O}_{DW}$
can be discriminated by using the lepton production channels, as
well as the channels involving quarks of the first two families, 
we concentrate in the present work on the process 
$e^-e^+ \to b \bar b$, which allows the study of the three
operators $\O_{qW}$, $\O_{qB}$, $\O_{bB}$. In the presence of
polarized beams, there are four possible 
observables that can be constructed, namely $\sigma_b$, $A^b_{FB}$,
$A^b_{LR}$ and $A^{pol(b)}_{FB}$, which allows to test and 
disentangle the three operators $\O_{qW}$, $\O_{qB}$, $\O_{bB}$.
The study therefore of the effects of the operators
($\overline{\O}_{DW}$, $\overline{\O}_{DB}$, $\O_{qW}$, 
$\O_{qB}$, $\O_{bB}$) can go beyond the usual treatment of 
NP effects, which
usually consists in taking one operator at a time, 
ignoring possible correlations.\par

In Section 2 we calculate the effects on $e^-e^+ \to f\bar f$
amplitudes starting by first using the equations of
motion, which   considerably simplifies the
computation. We apply this technique to all five operators, and
check that for $\overline{\O}_{DB}$ and $\overline{\O}_{DW}$  
it reproduces the results
obtained in a different way in ref.\cite{clean, cleanlong}. We also
establish the unitarity constraints for these five operators. They
allow to relate their coupling constants to the energy scale at which
the two-body scattering amplitudes saturate unitarity. At this energy,
new types of effects like the creation of heavy degrees of freedom,
should appear in order to restore unitarity. So this energy scale
corresponds to the effective NP scale $\lamNP$.
In Section 3 we compute the contributions to the observables
from the $\O_{qW}$, $\O_{qB}$ and $\O_{bB}$ operators,
using the $Z$-peak subtracted representation.\par

In Section 4 we apply the results to the LEP2 and NLC
energy ranges and we derive the observability limits for the
relevant NP couplings. Finally, Section 5 summarizes the physics 
issues of 
such an analysis, for what concerns the search for residual 
NP effects in $e^-e^+ \to f \bar f$ processes.\par

\section{$e^-e^+ \to b\bar b$ observables for studying 
$\overline{\O}_{DB}$, $\overline{\O}_{DW}$, $\O_{qW}$, 
$\O_{qB}$, $\O_{bB}$.}

As explained in the Introduction, the $Z$-peak subtracted
representation allows to disentangle the contributions from the
"derivative" operators leading to a strong $q^2$-dependence;
namely
$\overline{\O}_{DB}$, $\overline{\O}_{DW}$, $\O_{qW}$, 
$\O_{qB}$, $\O_{b}$. In this framework
the effect of the two bosonic operators has already 
been discussed in \cite{clean, cleanlong} 
using the results of \cite{Hag} obtained with the propagator
formalism. According to this, all NP contributions are expressed
as modifications to the $\gamma$ and $Z$ propagators. 
In this section we show
that the contribution of the above operators to 
$e^+e^-\to f\bar f$ can be more directly
computed in a very simple way using the equations of motion. \par
 
The effective Lagrangian describing the NP contribution from 
the $\O_i$ operators in (\ref{listDB}, \ref{listDW},
\ref{listqW}-\ref{listbB}), is
\bq  
\L_{eff} = \L_{SM} + \sum_i {f_i\over m^2_t} \O_i  \ \ .
\label{Leff} \eq
The SM equations of motion for 
the $B$ and $W$ fields
\bqa
D_\mu \overrightarrow W^{\mu\nu} & = & g 
\vec{J}^\nu ~ -~ i~\frac{g}{2} ~[D^\nu \Phi^\dagger
\overrightarrow \tau \Phi ~ -~ \Phi^\dagger
\overrightarrow \tau D^\nu \Phi ]\ \ , \label{listDmuW}\\[0.1cm] 
\partial_\mu  B^{\mu\nu} & = & g\prime 
J^\nu_{Y}~ -~ i~\frac{g\prime}{2}~ [D^\nu \Phi^\dagger \Phi ~ - 
~\Phi^\dagger  D^\nu \Phi ]\ \ , \ \label{listDmuB}
\eqa
with  $\vec{J}^\nu $ and
$J^\nu_{Y}$ being the  $SU(2)$ and hypercharge fermionic
currents respectively, allow us to express $\O_i$ as 
\bqa  
\overline{\O}_{DB} & =& 2 g'^2 J^\mu_{Y} J_{Y\mu}
+\frac{gg'^2}{\cw}~(v+H)^2 J^\mu_{Y}Z_{\mu}+
\frac{g^2g'^2}{8\cw^2}(v+H)^4 Z^\mu Z_\mu \ \ , 
\label{DBeqm} \\[0.1cm]
\overline{\O}_{DW} &= & 2 g^2 \vec{J}^\mu \vec{J}_\mu 
-\frac{g^3}{\sqrt{2}}~ (v+H)^2 (J^{+\mu} W_\mu^+
+J^{-\mu} W_\mu^-) -\frac{g^3}{\cw}~(v+H)^2 Z_\mu J^{3\mu}
\nonumber \\ 
&+& \frac{g^4}{8} ~(v+H)^4 \left (2 W^+_\mu W^{-\mu} +
\frac{1}{\cwd} Z^{\mu}Z_{\mu} \right ) \ \ , 
\label{DWeqm} \\[0.1cm]
\O_{qB} & = & -~g'(\bar q_L\gamma^{\mu}q_L) \cdot
\left (J_{Y\mu} +\frac{g}{4c_W}~(v+H)^2 Z_{\mu}\right ) \ 
\  , \label{qBeqm} \\
\O_{bB}& = & -~g'(\bar b_R \gamma^{\mu} b_R) \cdot
\left (J_{Y\mu} +\frac{g}{4c_W}~(v+H)^2 Z_{\mu}\right ) \ 
\  , \label{bBeqm} \\
&& \O_{qW}~ =~  -g \left (\bar q_L \frac{\vec{\tau}}{2} 
\gamma^{\mu}q_L \right ) \vec{J}_\mu + \nonumber \\
& &\frac{g^2}{4}\, (v+H)^2
\left \{\frac{1}{\sqrt{2}} (\bar q_L\gamma^\mu t^+ q_L)W^+_\mu
+ \frac{1}{\sqrt{2}} (\bar q_L\gamma^\mu t^- q_L)W^-_\mu
+\frac{1}{\cw}(\bar q_L\gamma^\mu \, \frac{\tau^3}{2}\, 
q_L) Z_\mu \right \}\ . \label{qWeqm}
\eqa
In (\ref{DBeqm}-\ref{qWeqm})  $\tau^i ~(i=1-3)$ are the usual three
Pauli matrices and $t^\pm \equiv (\tau^1 \pm i \tau^2)/2$, 
$q_L=(t_L, b_L)$ is the doublet of the left-handed quarks of the third
family, $J^\pm_\mu=J^1_\mu \pm i J^2_\mu$,
are the charged fermion currents,  and $W^\pm_\mu$ are
the fields absorbing $W^\pm$ respectively. \par

>From (\ref{DBeqm}-\ref{qWeqm}), note that these operators provide
$q^2$-independent NP contributions to
the $Zf\bar f$ couplings. These contributions are irrelevant for
our treatment though, since, together with any other NP
contributions from all other operators in \cite{Hag, topdyn},
they will be absorbed in the $Z$-peak observables used as
inputs. We also note from (\ref{DBeqm},\ref{DWeqm}), that 
$\overline{\O}_{DB}$ and  $\overline{\O}_{DW}$ induce tree level 
contributions to the $W$ and $Z$ masses, but no contribution to
the $\rho$ parameter measuring the neutral to ~charged current
ratio \cite{Hag}. Finally we also remark that there is no NP
contribution to the $\gamma f \bar f$ coupling.\par  

We next turn to the unitarity constraints on the couplings of
the $\O_i$ operators. Note that in (\ref{Leff}) we have normalized these
couplings to $m^2_t$. We apply the same techniques as in  
\cite{unit1, topdyn}. We consider the strongest unitarity
constraints arising from the two-body scattering amplitudes
and we identify the energy at which unitarity is saturated 
to the scale $\lamNP$. We find:\par

$\O_{qW}$: The most ~stringent  constraint for this operators
arises from the transitions among the $j=1$ colour-singlet
flavour-neutral channels 
$|t\bar t-+>$,  $|b\bar b-+>$, $|u \bar u-+>$,  $|c \bar c-+>$,
$|d \bar d -+>$,  $|s \bar s-+>$,
$|e\bar e-+>$,  $|\mu \bar \mu -+>$, $|\tau  \bar \tau-+>$,  
$|\nu_e \bar \nu_e-+>$, $|\nu_\mu \bar \nu_\mu -+>$,
$|\nu_\tau \bar \nu_\tau -+>$, $|ZHL>$, $|W^+W^-LL>$. The result
is
\bq
\label{fqW}
f_{qW} ~\simeq ~5.7\, \frac{\pi}{g}\, 
\left (\frac{m_t}{\lamNP} \right ) ^2 
~\simeq ~ 27.5\,  \left (\frac{m_t}{\lamNP} \right ) ^2\ \ .
\eq\par

$\O_{qB}$: The most stringent constraint also comes from the $j=1$
colour-singlet flavour-neutral channels
$|t\bar t-+>$,  $|b\bar b-+>$, $|u \bar u-+>$,  $|c \bar c-+>$,
$|d \bar d -+>$,  $|s \bar s-+>$,
$|t\bar t+->$,  $|b\bar b +->$, $|u \bar u+->$,  $|c \bar c+->$,
$|d \bar d +->$,  $|s \bar s +->$,
$|e\bar e-+>$,  $|\mu \bar \mu -+>$, $|\tau  \bar \tau-+>$,  
$|e\bar e+->$,  $|\mu \bar \mu +->$, $|\tau  \bar \tau +->$,
$|\nu_e \bar \nu_e-+>$, $|\nu_\mu \bar \nu_\mu -+>$,
$|\nu_\tau \bar \nu_\tau -+>$, $|ZHL>$, $|W^+W^-LL>$.
The result is
\bq
\label{fqB}
f_{qB} ~\simeq ~2.66 \, \frac{\pi}{g'}\, 
\left (\frac{m_t}{\lamNP} \right ) ^2 
~\simeq ~ 23.4 \,  \left (\frac{m_t}{\lamNP} \right ) ^2\ \ .
\eq\par

$\O_{bB}$: The most stringent constraint comes from the 
same $j=1$  channels as in the $\O_{qB}$ case.
The result is
\bq
\label{fbB}
f_{bB} ~\simeq ~3.45 \, \frac{\pi}{g'}\, 
\left (\frac{m_t}{\lamNP} \right ) ^2 
~\simeq ~ 30.4 \,  \left (\frac{m_t}{\lamNP} \right ) ^2\ \ .
\eq\par

$\ol{\O}_{DB}$: The most stringent constraint comes from 
$j=1$ channels which are singlet under colour, as well as 
under the horizontal $SU(3)$  group relating the three families.
Denoting by $U$ and $D$ the generic up and down quarks 
and by $E$ and $N$ the charged and neutral leptons, we write
these channels as $|U\bar U+->$, $|U \bar U-+>$, $|D \bar D+->$,
$|D\bar D-+>$, $|E \bar E +->$, $|E \bar E -+>$, $| N \bar N -+>$,
$|ZHL>$ and $|W^+W^-LL>$. After diagonalizing the relevant
$9\times 9$ transition matrix, we obtain the constraint 
\bq
\label{fDB}
f_{DB} ~\simeq ~0.56 \, \frac{\pi}{g'^2}\, 
\left (\frac{m_t}{\lamNP} \right ) ^2 
~\simeq ~ 13.8 \,  \left (\frac{m_t}{\lamNP} \right ) ^2\ \ .
\eq\par

$\ol{\O}_{DW}$: The situation is similar to the previous one,
but the channels are somewhat fewer now, namely $|U \bar U-+>$, 
$|D\bar D-+>$, $|E \bar E -+>$, $| N \bar N -+>$,
$|ZHL>$ and $|W^+W^-LL>$. The diagonalization of the transition
matrix gives 
\bq
\label{fDW}
f_{DW} ~\simeq ~  \frac{\pi}{g^2}\, 
\left (\frac{m_t}{\lamNP} \right ) ^2 
~\simeq ~ 7.36 \,  \left (\frac{m_t}{\lamNP} \right ) ^2\ \ .
\eq\par

When calculating process $e^-e^+ \to f \bar f$ 
$(f\not= t)$, at energies
higher than the $Z$ peak, all  fermion masses
can be neglected. In such a case, the only NP
contributions to which $e^-e^+ \to f \bar f$ is sensitive, are
those which can ~interfere with the SM ones and are 
therefore characterized by the fact that the 
antifermions $e^+$ and $\bar f$ have
helicities opposite to those of $e^-$ and $f$ respectively.
Thus in the helicity basis, the transition matrix is fully 
characterized by just the helicity of the outgoing $f$ and the
incoming $e^-$. Restricting to  $f\not= e$, the 
differential cross section for $e^-e^+ \to f \bar f$ is
then written as
\bqa 
\label{dsigmaeeff}
{d\sigma\over dcos\theta}& = & {\pi N_f\over 2q^2}  
\Bigg \{(1-P_e P'_e)
[(1+cos^2\theta)U_{11}+2cos\theta~ U_{12}] 
\nonumber \\
&+& (P'_e- P_e)
[(1+cos^2\theta)U_{21}+2cos\theta~ U_{22}]\Bigg \} \ ,  
\eqa
where $P_e~ (P_e^\prime)$ denote twice the average helicity of
the incoming $e^-~(e^+)$ beams, $N_f$ the QCD factor 
$N_f\simeq 3(1+{\alpha_s\over\pi}$) for quarks and $N_f=1$ 
for leptons, and
\bqa
U_{11} & =& \frac{1}{4}~[|F_{LL}|^2+|F_{RR}|^2+|F_{RL}|^2+
|F_{LR}|^2 ] \ \ ,\label{U11}  \\
U_{12} & =& \frac{1}{4}~[|F_{LL}|^2+|F_{RR}|^2-|F_{RL}|^2-
|F_{LR}|^2 ] \ \ ,\label{U12}  \\
U_{21} & =& \frac{1}{4}~[|F_{LL}|^2-|F_{RR}|^2+|F_{RL}|^2-
|F_{LR}|^2 ] \ \ ,\label{U21}  \\
U_{22} & =& \frac{1}{4}~[|F_{LL}|^2-|F_{RR}|^2-|F_{RL}|^2+
|F_{LR}|^2 ] \ \ .\label{U22} 
\eqa

The $F_{ij}$ in (\ref{U11}-\ref{U22}) denote the "reduced"
helicity amplitudes, where the angular dependence is removed.
The first index $i$
describes the helicity of the outgoing fermion $f$, while the second
index $j$ ~represents the helicity of the incoming $e^-$. 
As is seen from (\ref{dsigmaeeff}) and remarked
in \cite{Frere}, the angular dependence in the differential
cross section 
for fully polarized beams allows the complete separation of
all four independent $|F_{ij}|$ quantities.\par

Applying these for $f=b$, we have that the usual measurable 
quantities at any $q^2$; namely the integrated cross section, the 
forward-backward asymmetry, the longitudinal polarization 
asymmetry and the 
polarized forward-backward asymmetry, are obtained as
\bq
\label{observables}  
\sigma_b= {4\pi\over q^2}\left
(1+\frac{\alpha_s(q^2)}{\pi}\right ) U_{11}\ \ \ , \ \ \
 A^b_{FB} = {3\over4}{U_{12}\over U_{11}}\ \ \ , \ \ \ 
A^b_{LR} = {U_{21}\over U_{11}}\ \ \ , \ \ \
A^{pol,b}_{FB} = {3\over4}{U_{22}\over U_{11}} \   
\eq\par

As emphasized in \cite{Zsub}, in order to be able to take
into account any possible additional 
NP contribution not described by the operators in 
(\ref{DBeqm}-\ref{bBeqm}), we have to express the $Z$-peak
contribution to $F_{ij}$ directly in terms of the measured 
observables at LEP1/SLC.
Only then we are able to isolate the $q^2$-enhanced high energy
contribution induced by the ~aforementioned 5 operators. 
Apparently there are two
equivalent ways to do this. In \cite{Zsub, clean,
cleanlong} this was done by exactly absorbing all NP contributions,
either on or off the Z-peak, 
to modifications of the $\gamma$ and $Z$ propagators. 
The value of the Z-propagator on the Z-peak is then fixed by
experiment, and the NP induced by the five operators above, 
is completely described in terms of the
off-shell behaviour of the $\gamma-Z$ propagator.
This technique was then applied to the study of operators
$\O_{DB}$ and $\O_{DW}$. An alternative technique consists in
using the SM equations of motion, which naturally leads to a
representation where all NP is expressed either in 
terms Z-peak contributions, or in  
contact-term effects increasing with $q^2$. In the next Section
both techniques will be applied to study the operators  
$\O_{qW}$, $\O_{qB}$ and $\O_{bB}$. \par

\section{$Z$-peak subtracted forms for the $e^-e^+\to b\bar b$
Observables.}

The "reduced" helicity amplitudes $F_{ij}$ appearing
in (\ref{dsigmaeeff}-\ref{U22})  and ~describing 
 $e^-e^+ \to f \bar f$ with
$(e^-\mbox{-helicity} = j)$ and $(f\mbox{-helicity} = i)$, 
are written in the effective Born approximation as 
\bqa
F_{LL} & = & - \alpha(q^2) Q_f -\frac{3 \sqrt{\Gamma_f
\Gamma_e}}{\mz \sqrt{N_f}}\, \chi\, 
\frac{(1+\vtil_f)(1+\vtil_e)\epsilon(Q_f)}
{\sqrt{(1+\vtil_f^2)(1+\vtil_e^2)}}
\nonumber \\ 
&+&\frac{q^2}{ m^2_t}\, 
\Bigg [-\alpha(q^2)
\left (\frac{f_{DW}}{\swd}\tau^3_f + \frac{2
f_{DB}}{\cwd}Y_L^f\right )+
\delta_{bf}\, \sqrt{\frac{\alpha(q^2)}{4\pi}}
\left (-\,\frac{f_{qW}}{4\sw}+\frac{f_{qB}}{2\cw}
\right ) \Bigg ]\ , \label{FLL} \\
F_{RL} & = & - \alpha(q^2) Q_f +\frac{3 \sqrt{\Gamma_f
\Gamma_e}}{\mz \sqrt{N_f}}\, \chi\, 
\frac{(1-\vtil_f)(1+\vtil_e)\epsilon(Q_f)}
{\sqrt{(1+\vtil_f^2)(1+\vtil_e^2)}}
\nonumber \\ 
&+&\frac{q^2}{  m^2_t}\, 
\Bigg [-\alpha(q^2) \, \frac{2 f_{DB}}{\cwd}Y_R^f +
\delta_{bf} \sqrt{\frac{\alpha(q^2)}{4\pi}}
 \,\left (\frac{f_{bB}}{2\cw}\right ) \Bigg ]\ , 
\label{FRL}  \\ 
F_{LR} & = & - \alpha(q^2) Q_f +\frac{3 \sqrt{\Gamma_f
\Gamma_e}}{\mz \sqrt{N_f}}\, \chi\, 
\frac{(1+\vtil_f)(1-\vtil_e)\epsilon(Q_f)}
{\sqrt{(1+\vtil_f^2)(1+\vtil_e^2)}}
\nonumber \\ 
&+&\frac{q^2}{  m^2_t}\, 
\Bigg [-\alpha(q^2) \, \frac{4 f_{DB}}{\cwd}Y_L^f +
\delta_{bf} \sqrt{\frac{\alpha(q^2)}{4\pi}}
 \,\left (\frac{f_{qB}}{\cw} \right )\Bigg ]\ , 
\label{FLR}  \\
F_{RR} & = & - \alpha(q^2) Q_f -\frac{3 \sqrt{\Gamma_f
\Gamma_e}}{\mz \sqrt{N_f}}\, \chi\, 
\frac{(1-\vtil_f)(1-\vtil_e)\epsilon(Q_f)}
{\sqrt{(1+\vtil_f^2)(1+\vtil_e^2)}}
\nonumber \\ 
&+&\frac{q^2}{ m^2_t}\, 
\Bigg [-\alpha(q^2) \, \frac{4 f_{DB}}{\cwd}Y_R^f +
\delta_{bf} \sqrt{\frac{\alpha(q^2)}{4\pi}}
 \,\left (\frac{f_{bB}}{\cw}\right ) \Bigg ]\ . 
\label{FRR}
\eqa 
In (\ref{FLL}-\ref{FRR}),
\bq
\chi ~=~ \frac{q^2}{q^2 -\mzd +i \mz\Gamma_Z} \ \ ,
\eq
$\Gamma_f$ is the  $Z\to f \bar f$ partial width, and
\bq
\label{vtilf}
\vtil_f~=~ \frac{g_{Vf}}{g_{Af}}~=~1-4|Q_f|s_f^2 \ , 
\eq
where $s_f^2$ is the effective Weinberg
angle for the $f$-fermion. Thus $s_e^2$ (and $s_l^2$ assuming
lepton universality) is defined by the
longitudinal polarization asymmetry through
\bq
A_{LR}~=~\frac{2\vtil_e}{1+\vtil_e^2} \ ,
\eq
while $s_f^2$ is defined by the so called 
polarization forward-backward asymmetry
for the final $f$-fermion through \cite{cleanlong}
\bq
A_{f}~=~\frac{2\vtil_f}{1+\vtil_f^2} \ .
\eq
In the following we will apply this for $f=b$.
In practice, as we are only interested in first order manifestations of
the NP effects we can safely identify $\sw$ with $s_f$ or $s_e$
(or $s_l$) inside the coefficients multiplying these NP terms.\par

In the expressions for $F_{ij}$ given in (\ref{FLL} -\ref{FRR}),
the first term comes from the $\gamma$ exchange, the second 
from $Z$-exchange and the term proportional to $q^2$ arises from
the contact 4-fermion interactions induced by the $\O_i$ operators
we are studying. One way to apply the $Z$-subtraction technique
of \cite{Zsub} in these expressions is 
to neglect the photon and contact contributions for $q^2=\mzd$, 
and this way fix the coefficient of the
term proportional to $\chi$ using the LEP1/SLC measurements.
This is what was done in (\ref{FLL} -\ref{FRR}) and its validity
is based on the dominance of the $Z$-peak. Substituting
then these to (\ref{U11}-\ref{U22}), keeping only terms linear
in the NP couplings, we get through (\ref{observables}) the 
predictions for the four possible observables.\par

An alternative, and in principle more general  
way to ensure the correct $Z$-peak subtraction is 
the one suggested  in \cite{Zsub, clean, cleanlong} using the
results of \cite{Hag}. To ensure the correct $Z$-peak
subtraction, we first decompose the NP contact terms 
in (\ref{FLL}-\ref{FRR}) to a superposition of terms having the
photon and $Z$ Lorentz structures; i.e. $Q_f\gamma^{\mu}$ and 
$\gamma^{\mu}(g_{Vf}-g_{Af}\gamma_5)$. This way, we  
identify the $\gamma\gamma$, $ZZ$, $\gamma Z$,
$Z\gamma$ contributions to the neutral gauge boson propagator,
with the $Z$-peak subtracted
quantities $\Delta^{lf}\alpha(q^2)$, $R^{lf}(q^2)$, 
$V^{lf}_{\gamma Z}(q^2)$ and $V^{lf}_{Z\gamma}(q^2)$ defined in
\cite{Zsub}. 
This is done in a straightforward way, leading for 
$e^+e^-\to b\bar b$ (i.e. $f\equiv b$) to the results
\bq  
\Delta^{lb}\alpha(q^2)=
{q^2\over  m^2_t}\{4(c^2_W f_{DB}+s^2_W f_{DW})-{s^2_W\over g}f_{qW}
+{2s_Wc_W\over g}f_{qB}+{3\over g'}(1-2s^2_W/3)f_{bB}\} \,
\eq
\bq   
R^{lb}(q^2)={(q^2-M^2_Z)\over  m^2_t}\{-4(s^2_W f_{DB}
+c^2_W f_{DW})+{c^2_W\over g}f_{qW})
+{2s_Wc_W\over g'}(f_{qB}-f_{bB})\} \ , 
\eq
\bq
V^{lb}_{\gamma Z}(q^2)= {(q^2-M^2_Z)\over  m^2_t}\{4s_Wc_W
(f_{DB}-f_{DW}+{1\over4g}f_{qW})
-{2s_Wc_W\over g}(f_{qB}-f_{bB})\}\ 
\eq
\bq
V^{lb}_{Z\gamma}(q^2)=  {(q^2-M^2_Z)\over  m^2_t}\{4s_Wc_W
(f_{DB}-f_{DW}+{1\over4g}f_{qW})
+{2s^2_W\over g}f_{qB}+{3\over g}(1-2s^2_W/3)f_{bB}\}\ ,
\eq\par

Using then (\ref{dsigmaeeff}-\ref{U22}), we have that the 
general expression of the polarized $e^+e^-\to b\bar b$ angular
distribution is determined by
\bqa  
U_{11}=&&
{\alpha^2(0)\over9}[1+2\delta\tilde{\Delta}^{(lb)}\alpha(q^2)]
\nonumber\\
&&+2[{\alpha(0)\over3}]{q^2-M^2_Z\over
q^2((q^2-M^2_Z)^2+M^2_Z\Gamma^2_Z)}[{3\Gamma_l\over
M_Z}]^{1/2}[{3\Gamma_b\over N_b M_Z}]^{1/2}
{\tilde{v}_l \tilde{v}_b\over
(1+\tilde{v}^2_l)^{1/2}(1+\tilde{v}^2_b)^{1/2}}\nonumber\\
&&\times[1+
\tilde{\Delta}^{(lb)}\alpha(q^2) -R^{(lb)}(q^2)
-4s_lc_l
\{{1\over \tilde{v}_l}V^{(lb)}_{\gamma Z}(q^2)+{1\over 3\tilde{v}_b}
V^{(lb)}_{Z\gamma}(q^2)\}]\nonumber\\ 
&&+{[{3\Gamma_l\over
M_Z}][{3\Gamma_b\over N_b M_Z}]\over(q^2-M^2_Z)^2+M^2_Z\Gamma^2_Z}
\nonumber\\
&&\times[1-2R^{(lb)}(q^2)
-8s_lc_l\{{\tilde{v}_l\over1+\tilde{v}^2_l}V^{(lb)}_{\gamma
Z}(q^2)+{\tilde{v}_b\over3(1+\tilde{v}^2_b)}
V^{(lb)}_{Z\gamma}(q^2)\}]\ , 
\label{U11pro} \\[0.2cm]
U_{12}=&& 2[{\alpha(0)\over3}]{q^2-M^2_Z\over
q^2((q^2-M^2_Z)^2+M^2_Z\Gamma^2_Z)}
[{3\Gamma_l\over M_Z}]^{1/2}[{3\Gamma_b\over N_b
M_Z}]^{1/2}{1\over(1+\tilde{v}^2_l)^{1/2}(1+\tilde{v}^2_b)^{1/2}}
\nonumber\\
&&\times[1+
\tilde{\Delta}^{(lb)}\alpha(q^2) -R^{(lb)}(q^2)]\nonumber\\
&&+{[{3\Gamma_l\over
M_Z}][{3\Gamma_b\over N_b
M_Z}]\over(q^2-M^2_Z)^2+M^2_Z\Gamma^2_Z}
[{4\tilde{v}_l \tilde{v}_b\over(1+\tilde{v}^2_l)(1+\tilde{v}^2_b)}]
\nonumber\\
&&\times[1-2R^{(lb)}(q^2)-4s_lc_l
\{{1\over \tilde{v}_l}V^{(lb)}_{\gamma Z}(q^2)+{1\over 3\tilde{v}_b}
V^{(lb)}_{Z\gamma}(q^2)\}] \ , 
\label{U12pro} \\[0.2cm]
U_{21}=&& 2[{\alpha(0)\over3}]{q^2-M^2_Z\over
q^2((q^2-M^2_Z)^2+M^2_Z\Gamma^2_Z)}
[{3\Gamma_l\over
M_Z}]^{1/2}[{3\Gamma_b\over N_b
M_Z}]^{1/2}{\tilde{v}_b\over(1+\tilde{v}^2_l)^{1/2}
(1+\tilde{v}^2_b)^{1/2}}\nonumber\\
&&\times[1+\tilde{\Delta}^{(lb)}\alpha(q^2) -R^{(lb)}(q^2)
-{4s_lc_l\over3\tilde{v}_b}V^{(lb)}_{Z\gamma}(q^2)]\nonumber\\
&&+{[{3\Gamma_l\over
M_Z}][{3\Gamma_b\over N_b
M_Z}]\over(q^2-M^2_Z)^2+M^2_Z\Gamma^2_Z}
[{2\tilde{v}_l \over(1+\tilde{v}^2_l)}]\nonumber\\
&&\times[1-2R^{(lb)}(q^2)-4s_lc_l
\{{1\over \tilde{v}_l}V^{(lb)}_{\gamma
Z}(q^2)+{2\tilde{v}_b\over3(1+\tilde{v}^2_b)}
V^{(lb)}_{Z\gamma}(q^2)\}]  \ , 
\label{U21pro} \\[0.2cm]
U_{22}= && 2[{\alpha(0)\over3}]{q^2-M^2_Z\over
q^2((q^2-M^2_Z)^2+M^2_Z\Gamma^2_Z)}
[{3\Gamma_l\over
M_Z}]^{1/2}[{3\Gamma_b\over N_b
M_Z}]^{1/2}{\tilde{v}_l\over(1+\tilde{v}^2_l)^{1/2}
(1+\tilde{v}^2_b)^{1/2}}\nonumber\\
&&\times[1+\tilde{\Delta}^{(lb)}\alpha(q^2) -R^{(lb)}(q^2)
-{4s_lc_l\over \tilde{v}_l}V^{(lb)}_{\gamma Z}(q^2)]\nonumber\\
&&+{[{3\Gamma_l\over
M_Z}][{3\Gamma_b\over N_b
M_Z}]\over(q^2-M^2_Z)^2+M^2_Z\Gamma^2_Z}
[{2\tilde{v}_b \over(1+\tilde{v}^2_b)}]\nonumber\\
&&\times[1-2R^{(lb)}(q^2)-4s_lc_l
\{{2\tilde{v}_l\over(1+\tilde{v}^2_l)}
V^{(lb)}_{\gamma Z}(q^2)+{1\over
3\tilde{v}_b}V^{(lb)}_{Z\gamma}(q^2)\}] \ . \label{U22pro}
\eqa 
In practice, at the accuracy at which the NP effects can be observed, 
both ways of calculating $U_{ij}$ give identical
results. \par

\section{Observability limits at LEP2 and NLC}

The operators $\overline{\O}_{DB}$ and $\overline{\O}_{DW}$ 
contribute in a 
universal way to $e^+e^-\to f\bar f$ for any (light) fermion
$f$. The best constraints
on their couplings obviously come from the more accurate
precision measurements
obtained through lepton pair and light hadron production. 
This has been discussed in \cite{clean, cleanlong}. The
sensitivity limits on the couplings found in \cite{clean,
cleanlong} and the implied  
lower bounds on the associated unitarity scales are 
(compare (\ref{fDB}, \ref{fDW}))
\bq
\begin{array}{ccccc}
\frac{|f_{DB}|}{ m^2_t} ~(TeV^{-2})  & \lsim & 0.05 ,  
& 0.011 &  0.0056  \\   
\lamNP(DB) ~(TeV)    & \gsim & 17 &  35 & 50  
\label{sensDB}
\end{array}
\eq
\bq
\begin{array}{ccccc}
\frac{|f_{DW}|}{ m^2_t} ~(TeV^{-2})  & \lsim & 0.026 ,  
& 0.005 &  0.0028   \\   
\lamNP(DW) ~(TeV)    & \gsim & 17 &  38 & 51 
\label{sensDW}
\end{array}
\eq
for the LEP2, NLC500 (unpolarized) and  NLC500 (polarized) 
cases respectively, with a luminosity of $500fb^{-1}$ for
LEP2 and $20fb^{-1}$ for NLC.\par 

These bounds imply that the effect of the 
$\ol{\O}_{DB}$ ($\ol{\O}_{DW}$) operators on
the $b$ quark observables defined in (\ref{observables}),  
are  at LEP2 at most
\bqa   
|{\delta\sigma_b\over\sigma_b}| & \simeq & 0.0016~~~~(0.0054) 
\ , \\ 
|\delta A^b_{FB}| & \simeq & 0.0013~~~~(0.0008)  \ , \\
|\delta A^b_{LR}| & \simeq &  0.0040~~~~(0.0022) \ , \\
|\delta A^{pol,b}_{FB}| & \simeq &  0.0026~~~~(0.0012) \ ,
\eqa
while for NLC500(unpol,pol) they are at most
\bqa
|{\delta\sigma_b\over\sigma_b}| & \simeq & 
0.0031, 0.0016~~~~(0.0084, 0.0047)  \  , \\
|\delta A^b_{FB}| & \simeq &  0.0021, 0.0011~~~~(0.0010, 0.0006)
\ , \\
|\delta A^b_{LR}| & \simeq &  0.0071, 0.0036~~~~(0.0032, 0.0018) 
\ , \\
|\delta A^{pol,b}_{FB}| & \simeq & 0.0052, 0.0027 ~~~~(0.0023,
0.0013) \ .
\eqa\par

Thus, the effects of the $\ol{\O}_{DB}$ and $\ol{\O}_{DW}$
operators on the
$b\bar b$ observables,
expected on the basis of the sensitivity limits derived from the light 
fermion processes,
turn out to be much smaller than 
the expected experimental uncertainties listed in Table 1
for a $b\bar b$ tagging efficiency of 25\%.
As shown in this Table, these uncertainties appear to be
of the order of a few percent. So in the following
analysis of the $e^+e^- \to b\bar b$ process, working with 
the $Z$-peak subtracted
representation, we can ignore the uncertainties brought by the
$\overline{\O}_{DB}$ and $\overline{\O}_{DW}$ operators. 
Therefore, we restrict to
a 3-free parameter case involving $f_{qW}$, $f_{qB}$ and $f_{bB}$
only and proceed to the derivation of the observability limits.  
We write for each of the above four observables, $\A^i$, $(i=1,..4)$,
the  inequality 
\bq
\label{error}
|\sum_{j=1}^3 K^i_j f_j|~~ \geq ~~{\delta\A^i_{exp}\over\A^i_{SM}}
\ \ , \eq
where
\bq  K^i_j= {d((\A^i- \A^i_{SM})/\A^i_{SM})\over df_j}\ .
\eq
In (\ref{error}), $\delta\A^i_{exp}$ is assumed to be given by
the expected statistical uncertainty for the nominal collider
luminosity, assuming that the mean value of $\A_i$ is given by SM.
We then combine quadratically all such information coming from the $l$
available observables ($l=2$ at LEP2 and $l=2$ or $4$ at NLC). 
At one standard deviation this gives 
the observability domain which is outside the ellipsoid surface
\bq
\sum_{i=1}^l~~|\sum_{j=1}^3 [K^i_j f_j].
[{\delta\A^i_{exp}\over\A^i_{SM}}]^{-1}~|^2 ~~ = ~~1 \ .
\eq\par

The projections of this ellipsoid on the 3 planes spanned by pairs
of couplings $(f_j, f_k)$ are shown in Fig.1 . For the
unpolarized case, only two observables are available, namely 
$\sigma_b$ and $A^b_{FB}$. Thus, experiments provide only two
(linear) constraints on the system of the three NP couplings $f_j$.
The system is therefore not fully constrained, and the ellipsoid 
degenerates into bands, in (some) planes at
least. In Fig.1a,b,c, these bands are indicated with  {\it dotted }
lines for the unpolarized LEP2 case at 190GeV, and with {\it
solid } lines for the unpolarized NLC case at 500GeV. We should 
remark that in passing from the LEP2 case to the unpolarized NLC
one, an important reduction of the widths of the
bands occurs, from $O(1)$ to a few $10^{-2}~~ TeV^{-2}$. 
This is of course due to the strong
increase with $q^2$, of the contributions of the above
operators. \par

Finally if $e^{\pm}$ beam polarization is available, then two
more physical observables, namely $A_{LR}^b$ and
$A_{LR}^{pol,b}$ become possible, which transforms the band into
the ellipses shown in Fig.1a,b,c, and magnified in
Fig.2a,b,c.\par 

These results can be compared with a treatment
of the $\O_{qW}$, $\O_{qB}$ and
$\O_{bB}$ operators one by one, deriving the corresponding
sensitivities on the NP couplings and the related lower bounds
on the unitarity scales. Thus, for the LEP2, NLC(unpol),
and NLC(pol) cases, we get the respective results
\bq   
\begin{array}{ccccc}
\frac{f_{qW}}{m^2_t}~(TeV^{-2}) & \lsim & 0.60 & 0.036 & 0.032 \\
\lamNP ~ (TeV) & \gsim  & 6.8  & 27.6   & 29.3
\end{array}
\eq
\bq   
\begin{array}{ccccc}
\frac{f_{qB}}{m^2_t}~(TeV^{-2}) & \lsim & 0.41 & 0.030 & 0.018  \\
\lamNP ~ (TeV) & \gsim  & 7.6   &  27.9   &  36.0
\end{array}
\eq
\bq
\begin{array}{ccccc}
\frac{f_{bB}}{m^2_t}~ (TeV^{-2}) & \lsim &  0.58 &  0.030 & 0.013  \\
\lamNP ~ (TeV) & \gsim  &  7.3 & 31.8 & 48.4
\end{array}
\eq

At NLC, if polarization is available, one can observe that
the bounds obtained are less than a factor 2
stronger than those obtained in the 3-parameter case. This illustrates
the quality of the disentangling provided by the four observables. 

\section{Conclusions}

In this paper we have studied a special set of 
$dim=6$ $SU(3)\times SU(2)\times U(1)$ gauge invariant 
operators dubbed $\ol{\O}_{DB}$, $\ol{\O}_{DW}$, 
$\O_{qW}$, $\O_{qB}$ and $\O_{bB}$. These operators are
essentially characterized by the two
properties of being "non-blind" (i.e. affecting Z-peak
observables at tree level), and  of involving many 
derivatives which lead to strong energy ~dependencies.\par

We have emphasized that this second property allows to disentangle
the effects of these operators from all other ones ("blind" or
"non-blind"); provided the Z-peak subtraction technique is 
employed. This technique consists in fixing the values of the 
inputs using the LEP1/SLC
data, which then implies  that the high energy behaviour of 
$e^-e^+ \to f \bar f$ is only sensitive to the five operators just
mentioned.  We stress that this does not include any assumption on
the other 43 $dim=6$ $SU(3)\times SU(2)\times U(1)$ gauge
invariant and $CP$ symmetric operators \cite{Hag, topdyn},
since their effect is fully removed \cite{Zsub, clean,
cleanlong}.\par

We have then studied the observability of the effects of these five
operators at LEP2 and NLC. To appreciate the physical importance
of these, we have also established the unitarity constraints
for the two-body scattering amplitudes induced by the same
operators. This
allows to relate the coupling constant of each of them, 
to an effective oderator-depended NP scale; 
(defined as the energy at which new
degrees of freedom should be created in order to restore unitarity).
The observability limits can then be expressed as lower bounds for
these NP scales.\par

To disentangle the 5 operators above, we may proceed as follows.
In a first step, the processes $e^-e^+ \to f \bar f$, 
(where $f$ is a charged
lepton or a light  quark $u$, $d$, $s$, $c$), may be used to study the
possible appearance of the $\ol{\O}_{DW}$ and $\ol{\O}_{DB}$
operators. Once the situation concerning them is clarified,
the process $e^-e^+ \to b \bar b$ may be used to study the
remaining three operators $\O_{qW}$, $\O_{qB}$ and
$\O_{bB}$. We have shown that the uncertainties which may affect the
$\ol{\O}_{DW}$ and $\ol{\O}_{DB}$ effects
on the process $e^-e^+ \to b \bar b$
are weaker than the
experimental errors affecting the corresponding observables and thus
the determination of the $\O_{qW}$, $\O_{qB}$ and
$\O_{bB}$ effects.\par

In order to achieve the goal to fully study and discriminate
these 3 operators, it is mandatory to have polarized beams.
Because only then, we will 
have at least three independent $ b \bar b$ observables. This should
be possible at the NLC500 Collider.\par

The resulting observability limits presented in Fig.2abc, show a
possible determination of the $f_{qW}$, $f_{qB}$, $f_{bB}$ couplings
at the percent level, which means NP scales in the 20--30 TeV range. 
These bounds are just slightly weaker than those obtained, under the
same conditions, for the operators $\ol{\O}_{DW}$ and $\ol{\O}_{DB}$,
which were in the 50 TeV range.\par

At LEP2, where no polarization is available, we only have two
constraints affecting the contributions from the three operators
$\O_{qW}$, $\O_{qB}$ and $\O_{bB}$, which makes  their full
determination impossible. Non trivial constraints in the form of
bands for pairs of couplings are nevertheless obtained and shown
in Fig.1abc. The widths of these bands are of order 
$O(1~TeV^{-2}$).\par

For comparison we have also treated
the above operators one by one. In the NLC case the bounds obtained are
not much stronger than in the 3-parameter case. This illustrates
the quality of the disentangling provided by the four observables.
In the LEP2 case, the independent bounds for each coupling separately
are at the 0.4--0.6 $TeV^{-2}$ level, which means NP scales in 
the 6--8 TeV range.
This is comparable with what is expected for other operators and
better than what is obtained from a similar one by one treatment
at LEP1/SLC. This point was also emphasized in a recent paper
\cite{whisnant}.\par

In conclusion, we would like to reiterate on the
fact that, due to the lack of knowledge of the underlying dynamics,
NP manifestations can take many different forms (just note the
large number of possible $dim=6$ operators) and that it is therefore
essential to look for ways of disentangling the various classes
of effects.
An analysis of experimental
data at present and future colliders along the
lines presented in this paper, should bring significant information
in this direction, as it singles out a special class of
operators.


\begin{center}
{\bf Table 1: Expected accuracy on $e^-e^+\to b\bar b$ observables}\\
(for 25\% $b\bar b$ tagging efficiency.)\\
\vspace{0.5cm}
\begin{tabular}{|c|c|c|c|c|}
\hline
\multicolumn{1}{|c|}{Observable}&
\multicolumn{1}{|c|}{$\delta\sigma_b/\sigma_b$} & 
\multicolumn{1}{|c|}{$\delta A^b_{FB}$} & 
\multicolumn{1}{|c|}{$\delta A^b_{LR}$} &
\multicolumn{1}{|c|}{$\delta A^{pol,b}_{FB}$}
\\[0.1cm] \hline
$LEP2(190GeV,~ \L=500fb^{-1})$&$0.05$
&$0.04$  & $---$
&$---$ \\[0.1cm] \hline
$NLC(500GeV,~\L=20fb^{-1})$&$0.012$ &$0.01$ &$0.01$ &$0.01$ \\
[0.1cm] \hline
\end{tabular}
\end{center}

\newpage

\begin{center}

{\large \bf Figure captions}
\end{center}
\vspace{0.5cm}

{\bf Fig.1} Constraints from $e^+e^-\to b\bar b$ observables in the
3-free parameter case,\\ 
at LEP2 (without polarization) ({\it dotted }), ~at 
NLC (without polarization) ({\it solid }), ~at
NLC (with polarization) ({\it ellipse }).\\
{\bf(a)} projection on the ($f_{qB}$, $f_{bB}$) plane.\\
{\bf(b)} projection on the ($f_{qW}$, $f_{bB}$) plane.\\
{\bf(c)} projection on the ($f_{qW}$, $f_{qB}$) plane.\\

{\bf Fig.2} Constraints from $e^+e^-\to b\bar b$ observables in the
3-free parameter case, at
NLC with polarization.\\
{\bf(a)} projection on the ($f_{qB}$, $f_{bB}$) plane.\\
{\bf(b)} projection on the ($f_{qW}$, $f_{bB}$) plane.\\
{\bf(c)} projection on the ($f_{qW}$, $f_{qB}$) plane.\\

\newpage

\thispagestyle{empty}

\vspace*{-3.5cm}
\hspace*{-3cm}
\epsfig{file=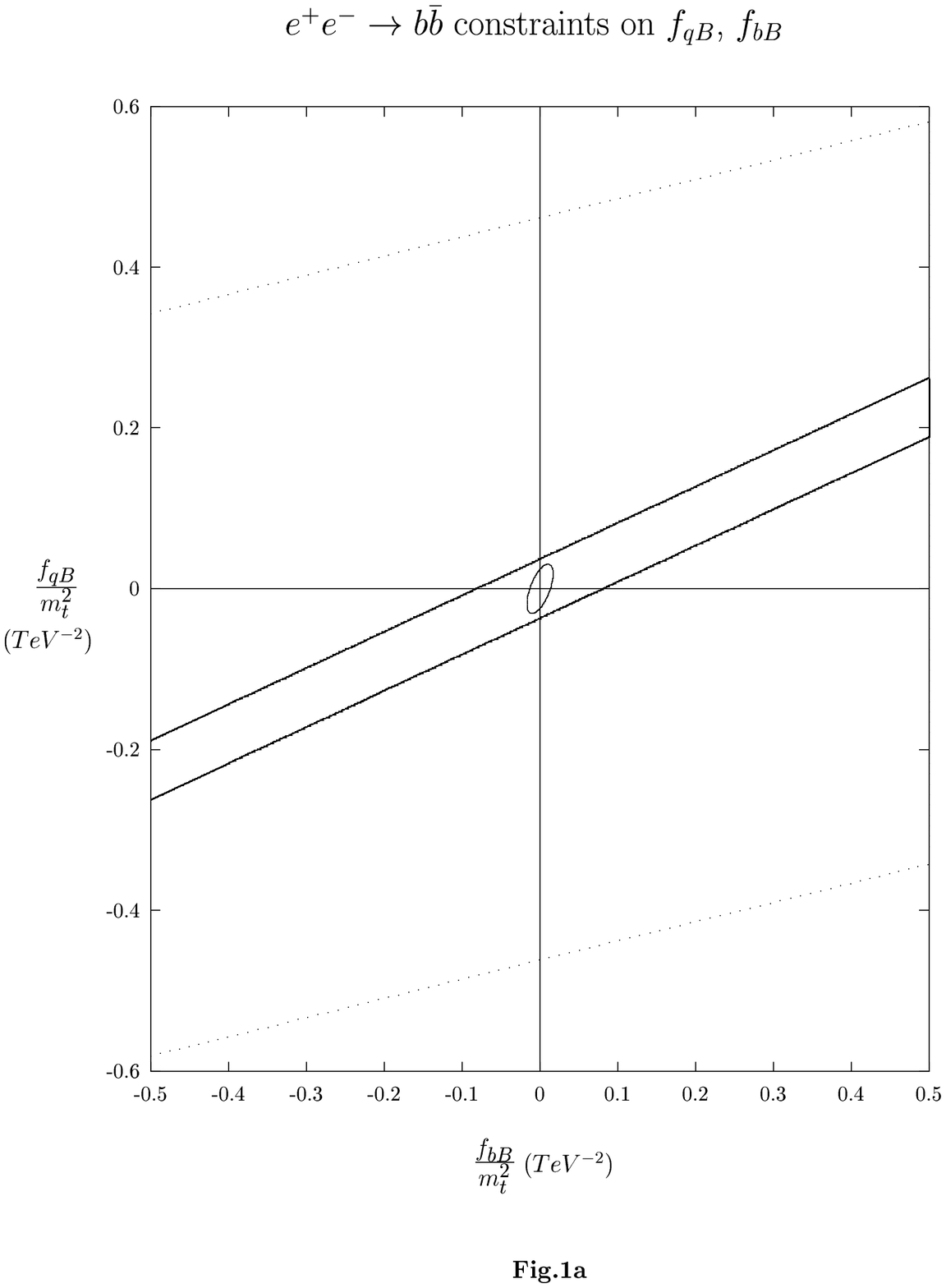}

\newpage
\thispagestyle{empty}

\vspace*{-3.5cm}
\hspace*{-3cm}
\epsfig{file=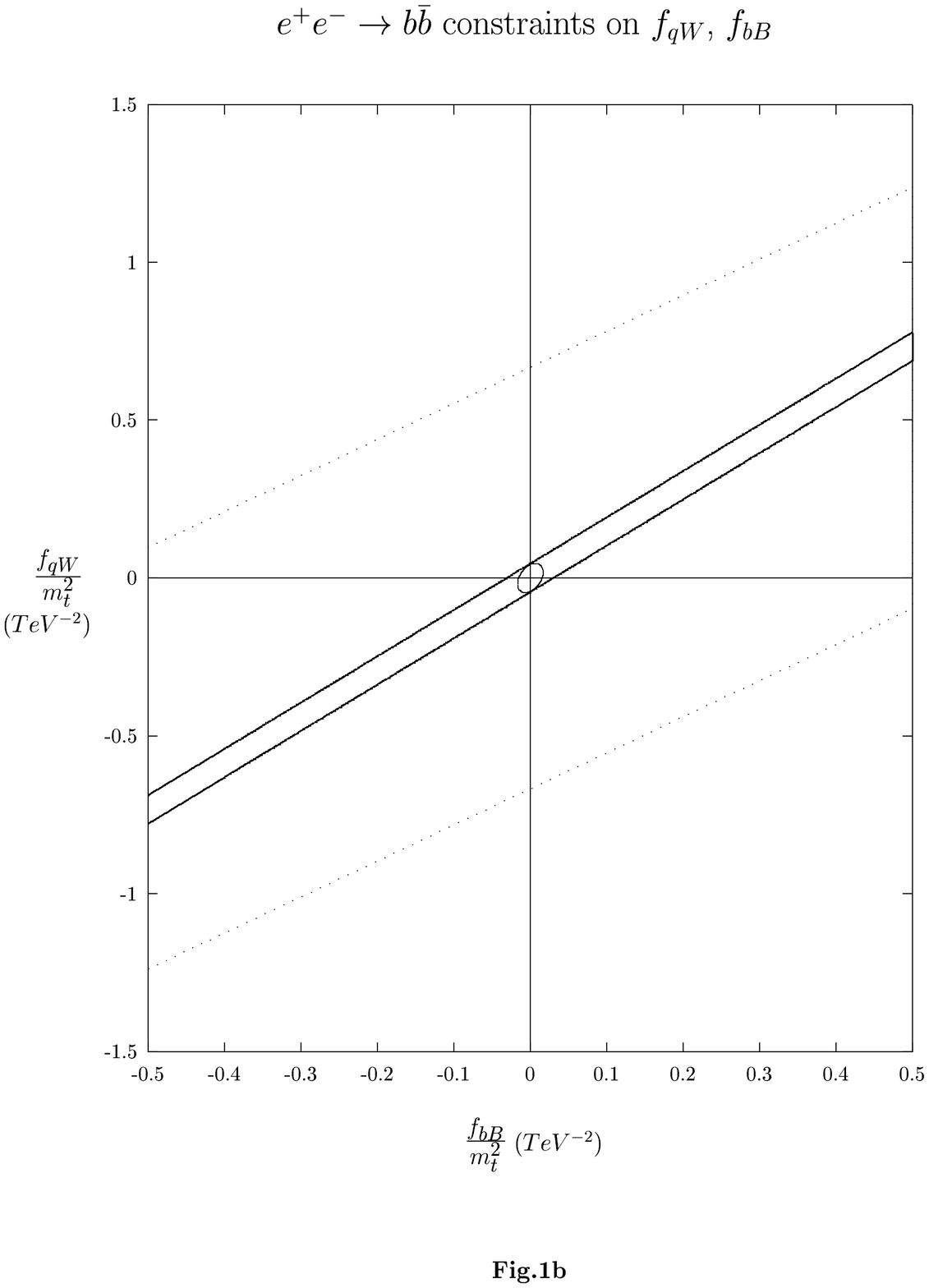}

\newpage
\thispagestyle{empty}

\vspace*{-3.5cm}
\hspace*{-3cm}
\epsfig{file=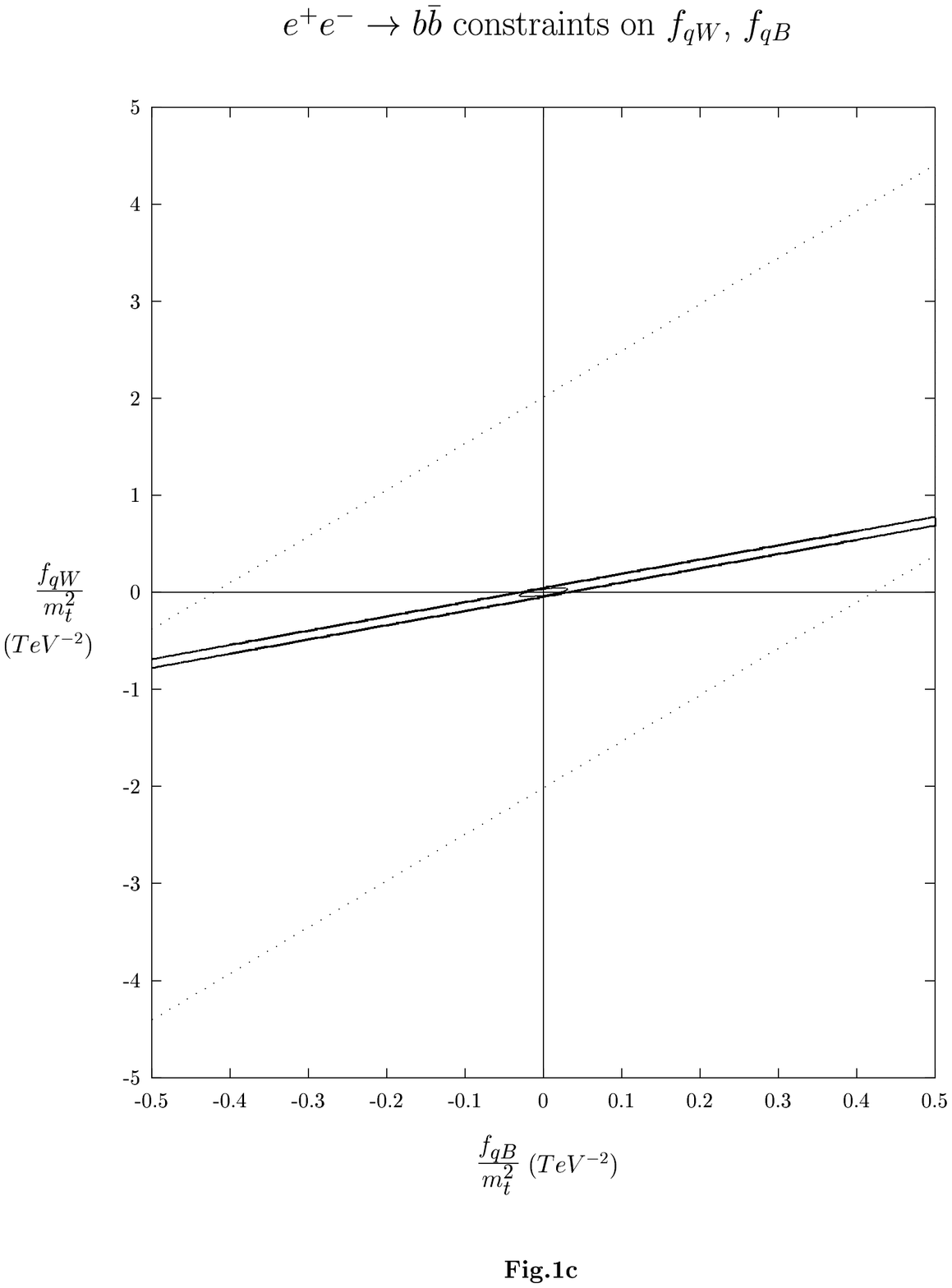}

\newpage
\thispagestyle{empty}

\vspace*{-3.5cm}
\hspace*{-3cm}
\epsfig{file=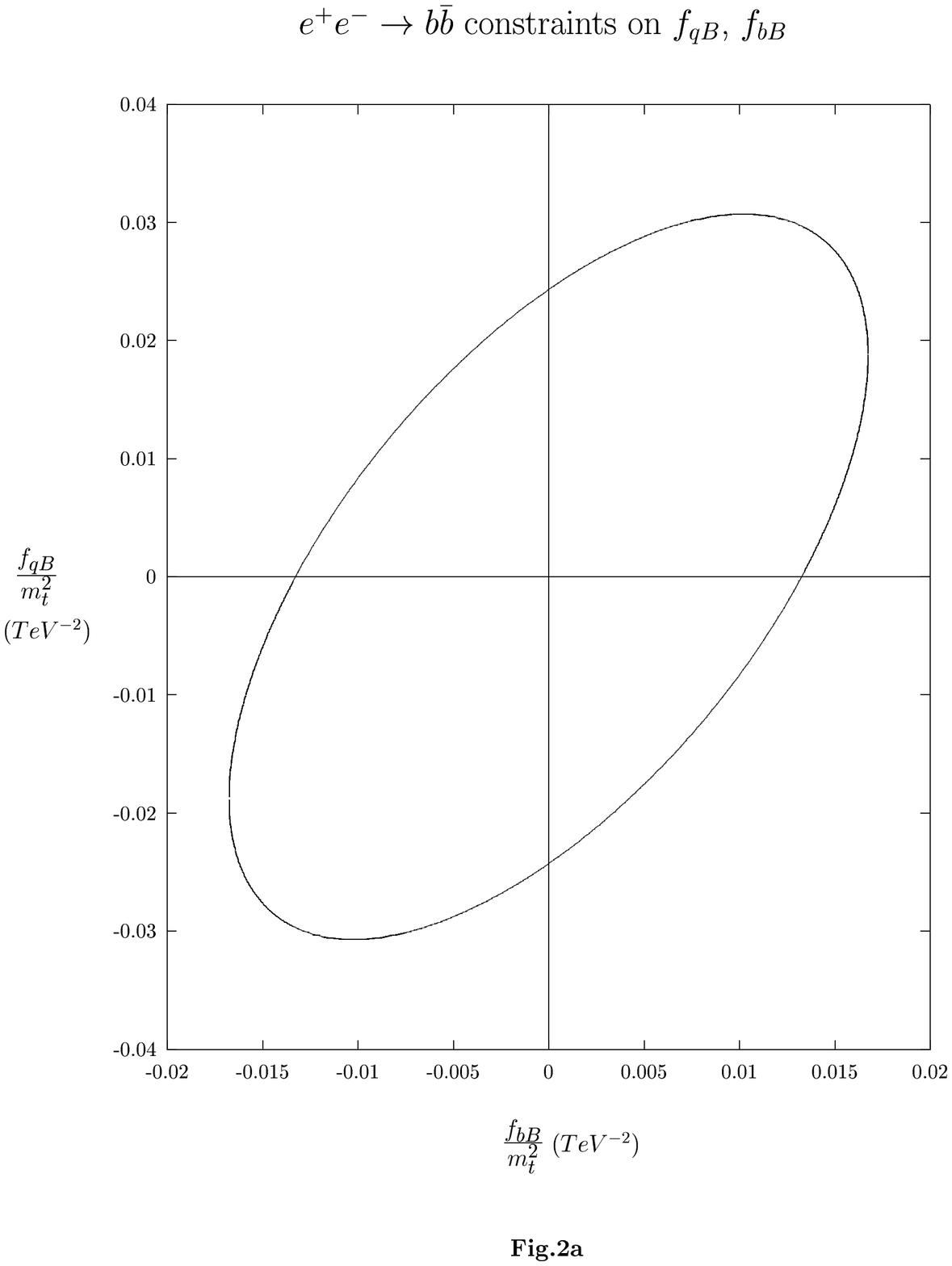}

\newpage
\thispagestyle{empty}

\vspace*{-3.5cm}
\hspace*{-3cm}
\epsfig{file=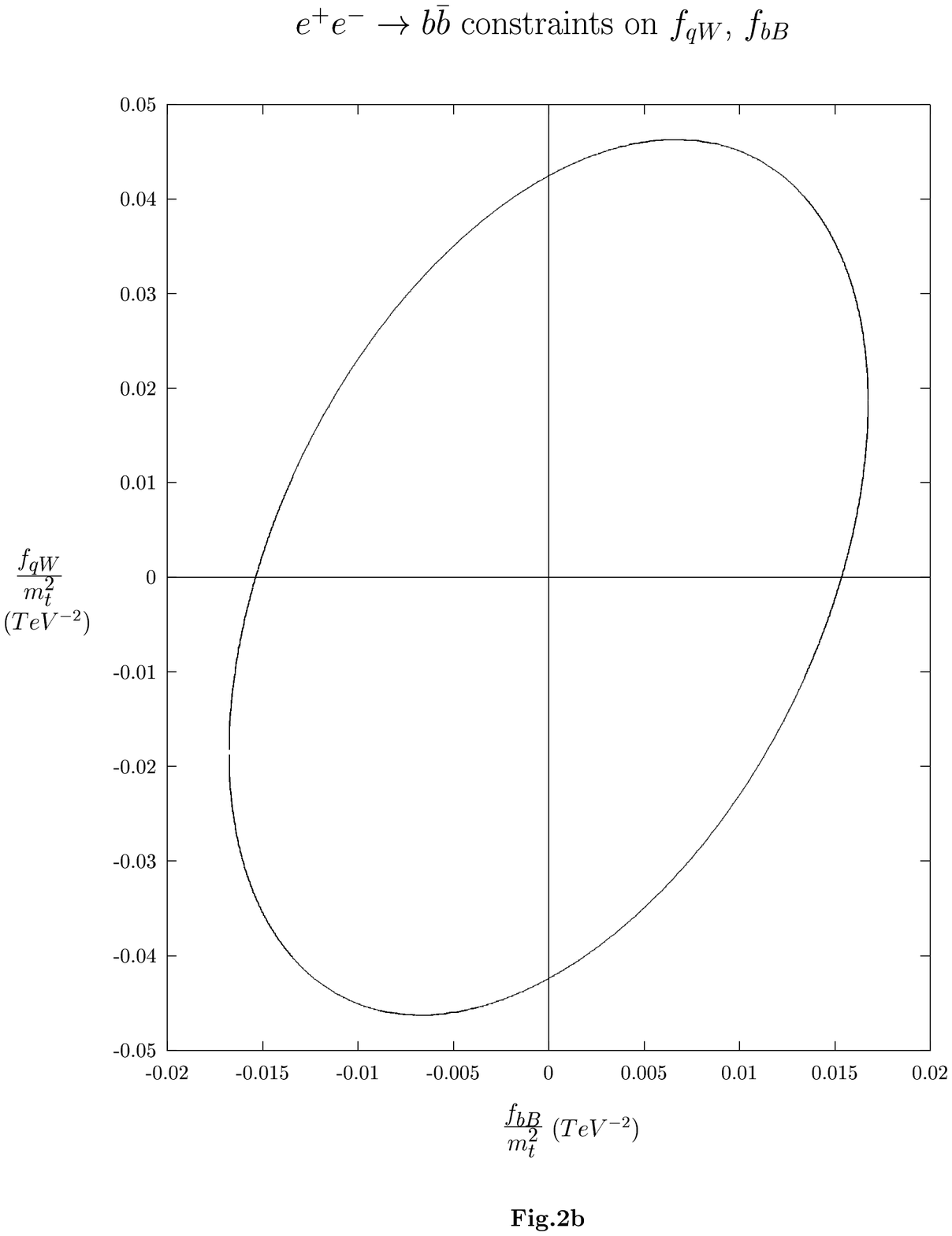}

\newpage

\thispagestyle{empty}
\vspace*{-3.5cm}
\hspace*{-3cm}
\epsfig{file=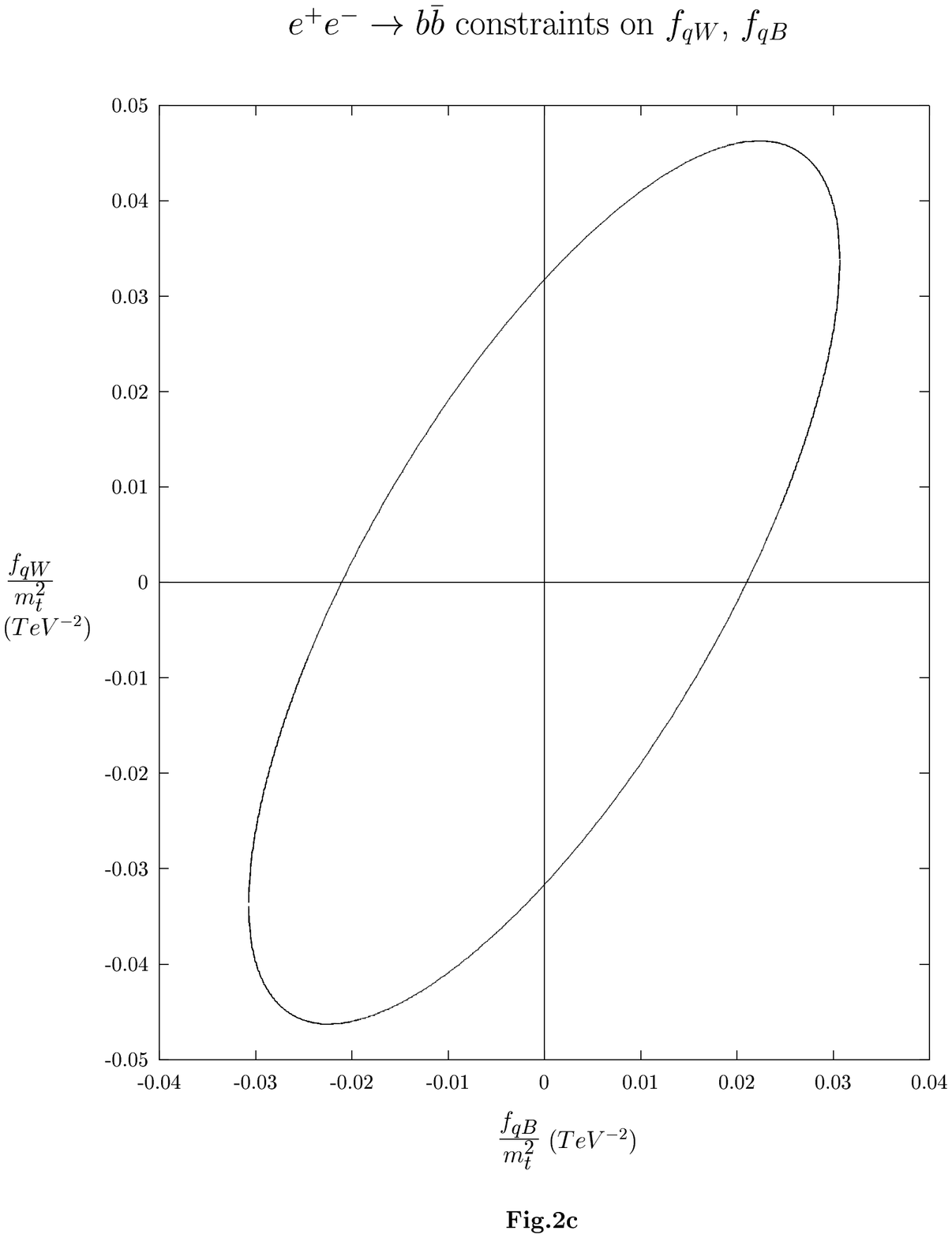}

\end{document}